\documentstyle[draft,aps,epsfig]{revtex}
\begin{document}
\draft
\title{Constraints on Hadronic Spectral Functions From Continuous \\
Families of Finite Energy Sum Rules}
\author{Kim Maltman\thanks{e-mail: maltman@fewbody.phys.yorku.ca}}
\address{Department of Mathematics and Statistics, York University, \\
          4700 Keele St., Toronto, Ontario, CANADA M3J 1P3 }
\maketitle
\begin{abstract}
Hadronic $\tau$ decay data
is used to study the reliability of various finite 
energy sum rules (FESR's).  For intermediate scales ($s_0\sim 2-3$
GeV$^2$), those
FESR's with
weights $s^k$ are found
to have significant errors, 
whereas those with weights having a zero
at the juncture ($s=s_0$) 
of the cut and circular part of the contour work very well.
It is also shown that a combination of two
such sum rules allows rather strong constraints to be placed
on the hadronic spectral function without sacrificing the excellent
agreement between OPE and hadronic representations.
The method then is applied to the strangeness $S=-1$
and pseudoscalar isovector channels, where it is shown to
provide novel constraints on those
ansatze for the unmeasured continuum parts of the spectral functions
employed in existing extractions of
$m_u+m_d$, $m_s$.
The continuum ans\"atz 
in the latter channel, in particular, is shown to produce a rather poor match
to the corresponding OPE representation, hence re-opening the question
of the value of the light quark mass combination $m_u+m_d$.
\end{abstract}

\section{Introduction}
As is well-known, for typical hadronic correlators $\Pi (s)$,
analyticity, unitarity, and the Cauchy theorem imply the
existence of dispersion relations which, due to
asymptotic freedom, allow non-trivial input in the form of
the operator product expansion (OPE) at large spacelike $s=q^2=-Q^2$.
The utility of these relations can be
improved by 
Borel transformation\cite{svz,rry,narisonbook}, 
which introduces both an exponential weight,
$exp(-s/M^2)$, (where
$M$, the Borel mass, is a parameter of the transformation),
on the hadronic (spectral integral) side 
and a factorial 
suppression of contributions from higher dimension operators on
the OPE side.  The exponentially decreasing weight allows
rather crude approximations for the large $s$ part of the spectral
function to be tolerated.  Typically, one employs essentially
a ``local duality'' approximation, i.e., uses 
the OPE version of the spectral function 
for all $s$ greater than some ``continuum threshold''.
Competition between optimizing
suppression of contributions from the crudely modelled ``continuum''
and
convergence of the OPE 
usually results in a
``stability window'' in $M$ 
for which neither contributions
from the continuum, nor those from the highest
dimension operators retained on the OPE side, are completely
negligible.  The resulting uncertainties
have to be carefully monitored 
to determine the reliability of a given
analysis\cite{leinweber}.
The presence of the decreasing exponential weight 
also means that the method is less sensitive
to the parameters of the higher resonances in the channel in question.

In this paper we investigate the alternative to Borel-transformed
(SVZ) sum rules provided by those finite energy sum rules 
(FESR's)\cite{shankar,mpr77,ck7778,fndr79,bldr85} generated by
integration over the ``Pac-man''
contour (running from $s_0$ to threshold
below the cut on the real timelike axis, back from threshold to $s_0$
above the cut, and closed by a circle of radius $s_0$ in the complex
$s$ plane).  One advantage of such FESR's is the absence of an 
exponentially decreasing weight.  Indeed, if
$\Pi (s)$ is a hadronic correlator without kinematic
singularities,
and $w(s)$ any analytic weight function, then, with the spectral
function $\rho (s)$ defined as usual,
\begin{equation}
\int_{s_{th}}^{s_0}\, w(s) \rho (s)\, ds = -{\frac{1}{2\pi i}}
\int_{\vert s\vert =s_0}\, w(s)\Pi (s)\, ds \ .\label{fesrbasic}
\end{equation}
Such FESR's have to date usually employed
integer power weights, $w(s)=s^k$, $k=0,1,2$ (see, {\it e.g.},
the recent extraction of $m_u+m_d$ in Ref.~\cite{bpr}
using the isovector pseudoscalar correlator),
but are, of course, valid for
any $w(s)$ analytic in the region of the contour.  

One interesting FESR involving a non-integer-power weight
is that relevant to 
hadronic $\tau$ decay.  
Neglecting the tiny 
contributions proportional to $(m_d-m_u)^2$ in the
isovector vector (IV) current correlator 
(which are, in any case, 
hard to handle reliably on the OPE side\cite{kmtau,pp98}),
the ratio of the non-strange hadronic to electronic
widths is proportional to
\begin{equation}
\int^{m_\tau^2}_{4m_\pi^2} {\frac{ds} {m_\tau^2 }} \,
\left( 1-{\frac{s}{m_\tau^2}}\right)^2 
\left( 1 + 2 {\frac{s}{m_\tau^2}}\right) 
\rho_\tau^{(0+1)}(s)  
\label{tauhadrep}\end{equation}
with $\rho_\tau^{(0+1)}(s)$ the sum of longitudinal and transverse
contributions to the corresponding spectral function\cite{taurefs}.  
This is the
hadronic side of a FESR with weight 
\begin{equation}
w_\tau (s)={\frac{1} {m_\tau^2 }} \,
\left( 1-{\frac{s}{m_\tau^2}}\right)^2 
\left( 1 + 2 {\frac{s}{m_\tau^2}}\right) \ ,\label{tauweight}
\end{equation}
the OPE side of which is
\begin{equation}
{\frac{i}{2\pi}} \oint_{|s|=m_\tau^2} {\frac{ds}{ m_\tau^2}}\,
\left( 1- {\frac{s}{ m_\tau^2}}\right)^2  \left( 
1 + 2 {\frac{s}{m_\tau^2}}\right) \Pi_{V,ud}^{(0+1)}(s)\ .
\label{OPErep}\end{equation}
Eq.~(\ref{OPErep}) thus provides an expression for the 
non-strange 
hadronic $\tau$ decay width\cite{tauope,dp92,pichrev} which requires as input
only $a(m_\tau^2)=\alpha_s(m_\tau^2)/\pi$
and the relevant $D=4$ and $D=6$ condensates (these latter,
in fact, give rather
small contributions)\cite{tauope,dp92,pichrev}.
This representation works extremely well, in the
sense that the $a(m_\tau^2)$ value required by $\tau$ decay data
(see Refs.~\cite{ALEPHalphas,ALEPH} and earlier references cited therein)
is nicely consistent, after running, with that
measured experimentally at the $Z$ mass scale\cite{pichrev}.
One can also verify that the success of the
underlying FESR is not a numerical accident 
by comparing the
hadronic and OPE sides for a
range of $s_0$ values $<m_\tau^2$\cite{ALEPH98}.  As shown by
ALEPH, and reiterated below, 
the agreement between the two representations is excellent
for all $s_0$ between $2\ {\rm GeV}^2$ and
$m_\tau^2$.

The success of the $\tau$ decay FESR has a simple physical explanation.
As argued in Ref.~\cite{pqw}, for large enough $s_0$, 
the OPE should provide a good representation
of $\Pi (s)$ over most of the circle $\vert s\vert =s_0$.
When local duality is not yet valid, however, 
this representation will necessarily break down over some region near the
timelike real axis.
Since, with $\Delta$ some typical hadronic scale,
the problematic region represents a fraction $\Delta /\pi s_0$ of the
full circle, one might expect the error on
the OPE side of a given FESR to be
$\sim\Delta /\pi s_0$.  Consider, however, a correlator having
perturbative contribution of the form $Q^{2n}\left(
1+c_1 a(Q^2)+c_2 a(Q^2)^2 + \cdots\right)$, with $n$ positive.
Expanding this expression in terms of $a(s_0)$, one
obtains $Q^{2n}\left( 1+c_1 a(s_0)+
{\cal O}\left( a(s_0)\right)^2\right)$, 
where the coefficient of the second order
terms now involves $log (Q^2/s_0)$ (for details, see e.g. 
Ref.~\cite{kniehl96}).  Integrating around $\vert s\vert =s_0$,
for any analytic $w(s)$,
the first surviving contribution is then the
${\cal O}\left( a(s_0)\right)^2$ logarithmic term.  This logarithm,
associated with the perturbative representation of $\alpha_s(Q^2)$,
has maximum modulus precisely
in the region (on either side of the cut on the time-like real axis)
for which the perturbative representation is least reliable. 
FESR's associated with weight functions (such as
$w(s)=s^k$) not suppressed near
$s=s_0$ can thus have errors potentially much greater than those suggested
by the naive estimate above.  We illustrate this point for the
case of the IV correlator below.  For
hadronic $\tau$ decay, however, phase space naturally produces
a (double) zero of $w_\tau (s)$
at $s=m_\tau^2$, and
this suppression of contributions from the region of the contour near
the real timelike axis results in
a very accurate FESR.
We will see, in Section II, that other weight functions with zeros
at $s=s_0$ also produce very reliable FESR's in the IV channel.
We will then use such weights, and corresponding FESR's, in
Section III, to 
studying the pseudoscalar isovector (PI)
and strangeness $S=-1$ scalar (SS) channels
(relevant to the extraction of the light quark mass combinations 
$m_u+m_d$\cite{bpr,pnew,newqm}
and $m_s+m_u$\cite{jm,cps,cfnp,newqm,kmtau,kmms}).

\section{Lessons from hadronic $\tau$ decay}
As noted above, FESR's involving weights, $w(s)$, with
$w(s_0)\not= 0$ have significant {\it potential} uncertainties
if local duality
is not yet valid at scale $s_0$.  To quantify this statement, consider
the $s^k$-weighted FESR's for the IV 
channel.  
In Table~\ref{Table1}
we list, as a function of $s_0$, the hadronic ($I_k^{ex}$) 
and OPE ($I_k^{OPE}$) sides of these sum rules,
\begin{eqnarray}
I_k^{ex}&\equiv&
\int_{4m_\pi^2}^{s_0}\, s^k \rho^{ex} (s)\, ds \ , \\
\label{skaleph}
I_k^{(OPE)}&\equiv&-{\frac{1}{2\pi i}}
\int_{\vert s\vert =s_0}\, s^k\Pi_{V,ud}^{(0+1)}(s)\, ds \ ,
\label{skope}
\end{eqnarray}
for $k=0,1,2,3$.
The hadronic side is evaluated using the spectral
function, $\rho^{ex}(s)$, measured by ALEPH\cite{ALEPH},
while for the OPE side 
we employ the known OPE for $\Pi_{V,ud}^{(0+1)}(s)$\cite{tauope,pichrev}, 
together with
(1) ALEPH values for $a(m_\tau^2)$ and the $D=6$ condensate terms (from the
non-strange decay analysis alone), (2) the
gluon condensate of Ref.~\cite{narison97},
(3) the GMO relation, $< 2m_\ell \bar{\ell}\ell >=-m_\pi^2f_\pi^2$,
(4) quark mass ratios from Chiral Perturbation Theory 
(ChPT)\cite{leutwylerqm}, 
(5) $0.7<\,\left[ <\bar{s}s>/<\bar{\ell}\ell >\right]\,< 1$,
as in Refs.~\cite{jm,cps,cfnp},
and (6) four loop running,
with contour improvement\cite{dp92}, for the perturbative contributions.

As seen from the Table, the errors in the
$s^k$-weighted FESR's
are significant, except near $s_0\sim 2.8\ {\rm GeV}^2$,
where the hadronic and OPE representations happen to cross.  The
worsening of
agreement for
$s_0$ above $\sim 2.8$ GeV$^2$ simply
reflects the facts that (1) local duality is not valid for
$s_0\sim 3$ GeV$^2$ and (2) the problematic region of the circular
part of the contour contributes
significantly to the OPE side of the sum rule for $w(s)=s^k$,
as one would expect.  Note that the
situation cannot be improved by taking ``duality ratios''
(ratios of such sum rules corresponding to different values of $k$)
since, as pointed out in Ref.~\cite{bgm}, 
if one insists
on a match between the hadronic and OPE versions of such a duality
ratio for $s_0$ values lying in some ``duality window'',
then the spectral function 
is constrained to match (up to an undetermined overall multiplicative constant)
that implied by the OPE 
for all $s_0$ in that window. 
If $s_0$ lies in the region of validity of local duality, this is
not a problem, but if it does not
(for example, if distinct
resonances are still present), 
then contributions from the problematic part of the
contour cannot have fully cancelled in the ratio.

The situation is much improved if we consider FESR's corresponding
to weights with a zero at $s=s_0$.  For
reference we give, in Table II, the experimental
(hadronic) and OPE sides of the FESR for the (double zero) combination
$I_0-3I_2+2I_3$ relevant to hadronic $\tau$ decay (see also
the discussion in Ref.~\cite{ALEPH98}).  As noted 
earlier, the match between the two sides
is very good, even at low scales.  We consider also the FESR's corresponding
to Eq.~(\ref{fesrbasic}), 
based on the weights
$w_{k,k+1}(s)=\left(s/s_0\right)^k\, -\, \left(s/s_0\right)^{k+1}$
which have only a simple zero at $s=s_0$.
Denoting the hadronic and OPE sides
by $J_{k,k+1}^{(ex,OPE)}(s_0)$, we have
\begin{equation}
J_{k,k+1}^{(ex,OPE)} (s_0)={\frac{1}{s_0^k}}
I_k^{(ex,OPE)} (s_0)-{\frac{1}{s_0^{k+1}}}I_{k+1}^{(ex,OPE)} (s_0)\ .
\label{singlepinch}
\end{equation}
The results for the cases $k=0$ and
$k=1$ are given in Table II.  Evidently,
even a simple 
zero at $s=s_0$ is enough the suppress
contributions from the problematic part of the contour
sufficiently to produce sum rules that are very reliable, again
even down to rather low scales.

The constraints on the hadronic spectral function obtained by combining
the $w_{01}$ and $w_{12}$ sum rules 
are actually rather strong.
To see this, note that a 
general linear combination of the two sum rules involves, for some
constant $A$,
a weight function proportional to
\begin{equation}
w(A,s) = \left( 1-{\frac{s}{s_0}}\right)\left( 1+A{\frac{s}{s_0}}\right)\ .
\label{aweight}
\end{equation}
For $A<-1$, this weight has a second zero in the hadronic integration region 
which moves to lower $s$ as $A$
is decreased.  To the left of this zero, the
spectral function is weighted positively, to the right,
negatively.  Dialing the crossover location
(by varying $A$) then places rather strong
constraints on the spectral function, provided that (1)
the OPE representation remains accurate over the range of
$A$, $s_0$ values employed and (2) 
the errors on the OPE side are not unduly
exacerbated by cancellations in forming the combination of
the two sum rules.
For the IV channel, it is
straightforward to demonstrate that
the errors are, indeed, not amplified, and 
that the resulting OPE and hadronic
representations do, indeed, remain in excellent agreement, for
$-9\leq A\leq 9$ (over which range the weight
function varies from having no second zero to having one
below the $\rho$ peak) and for a range
of $s_0$ values extending well below $m_\tau^2$.  We do not
display these facts explicitly since
the central values for the OPE and hadronic representations
follow from those already
given in the Table.
It is also worth stressing that, not only does the OPE representation
match very well with the hadronic one based on experimental data,
but that making a sum-of-resonances ans\"atz for the spectral function
and fitting its parameters to the OPE representation produces a model
spectral function in good agreement with the experimental one (for example,
the resulting $\rho$ decay constant differs from the experimental by
less than the experimental error).

\section{Applications to other channels}
We now consider FESR's for 
two channels of relevance to the 
extraction of the light quark masses.  For the
PI channel, unmeasured continuum contributions
to the spectral function contribute roughly three-quarters of
the hadronic side of the (integer power weighted) FESR which
determines $(m_u+m_d)^2$, while for the SS channel,
experimental constraints exist only on the $K\pi$ portion of
the spectral function.  We employ FESR's based on the weights
$w(A,s)$ in
both these channels;
in the former, to test the plausibility of the ans\"atz employed
for the continuum part of the spectral function\cite{bpr,pnew} and, in the
latter, to test the viability of certain assumptions/approximations 
made in the earlier analyses.

For the PI channel,
$m_u+m_d$ is extracted from
sum rules for the correlator, $\Pi_5(q^2)=i\int d^4x\, e^{iq\cdot x}
<0\vert \left(\partial_\mu A^\mu (x)\partial_\mu A^\mu (0)\right)\vert 0> $,
where $A^\mu$ is the isovector axial vector current.  With $\rho_5(s)$
the corresponding spectral function, one has, using
$w(s)=s^k$\cite{bpr,pnew},
\begin{equation}
\int_0^{s_0}\, ds\, s^k\, \rho_5(s)\, =\,
{\frac{3}{8\pi^2}}\left[ m_u(s_0)+m_d(s_0)\right]^2
{\frac{s_0^{k+2}}{k+2}}\left[ 1+R_{k+1}(s_0)+D_k(s_0)\right]
+\delta_{k,-1}\Pi_5(0)
\label{bprsr}
\end{equation}
where $R_{k+1}(s_0)$ (the notation is that of Ref.~\cite{bpr})
contain the perturbative corrections, 
$D_k$ the contributions from higher dimension operators,
and $\Pi_5(0)$ is
determined by $f_\pi$, $m_\pi$ and the combination $2L_8^r-H_2^r$
of fourth order ChPT low energy constants (LEC's)
(see Ref.~\cite{gl85}).  The analysis of Refs.~\cite{bpr,pnew}
(BPR) proceeds by (1) adjusting
the relative strength of $\pi (1300)$ and $\pi(1800)$
contributions to a continuum spectral ansatz
using the $k=0$ to $k=1$ duality ratio, (2) fixing
the overall scale of this ans\"atz by normalizing
the sum of the resonance tails to 
the leading order ChPT expression for $\rho_5(s)$
at continuum threshold,
(3) (with $\rho_5(s)$ so fixed)
using the $k=0$ and $k=-1$ sum rules to extract $m_u+m_d$ and
$2L_8^r-H_2^r$, respectively.
A number of possible problems exist with this analysis.  
First (see Ref.~\cite{bgm})
there are potential
dangers in the overall normalization prescription, associated with
the fact that continuum
threshold is rather far from the resonance peak locations.
Second, the value of the LEC combination obtained implies an
unusual value for the light quark condensate ratios. 
(The combination $2L_8^r-H_2^r$ is related to $2L_8^r+H_2^r$, 
which controls flavor breaking in the condensate ratios\cite{gl85}.  
With standard values for $L_8^r$\cite{gl85}, the BPR 
$2L_8^r-H_2^r$ value corresponds to $< \bar{s}s>/< \bar{u}u>=1.30\pm 0.33$.)
Finally, the presence of the $\pi (1800)$ signals that
one is 
not in the region of local duality, and hence that
non-negligible errors may be present in
$s^k$-weighted FESR's (and duality ratios thereof).  We can
investigate this latter question by considering those
additional constraints on
the BPR continuum spectral ans\"atz obtained from the $w(A,s)$
family of FESR's.
As input to the OPE side we use the latest ALEPH
determination of $a(m_\tau^2)$, the condensate
values employed in Refs.~\cite{bpr,pnew}, the most recent
value of Ref.~\cite{pnew} for $m_u+m_d$, and the four-loop 
contour-improved version
of the perturbative contributions.  Apart
from the small decrease in the ALEPH value of $a(m_\tau^2)$
between 1997 and 1998 the input to the OPE analysis is, therefore,
identical to that of Refs.~\cite{bpr,pnew}.
To be specific in tabulating results, we have employed,
on the hadronic side, the updated continuum ans\"atz of 
Ref.~\cite{pnew}
(the situation is not improved if one uses instead any of the
earlier ansatze of Ref.~\cite{bpr}). 
In Table III we present the hadronic (had)
and OPE sides of the resulting sum rules 
for 
$s_0$ in the BPR duality window and the range
$0\leq A\leq 9$.  Note that the best duality match from
Ref.~\cite{pnew} corresponds to $s_0=2.0\sim 2.4$
GeV$^2$.
In all cases, the known $\pi$ pole
contribution has been 
subtracted from both sides of the sum rule; the results thus provide
a direct test of the continuum spectral ans\"atz.
While
one cannot guarantee that the $w(A,s)$ sum rules
will work as well in the PI as in the IV channel,
there are clear physical grounds for expecting them to be more reliable
than those based
on the weights $w(s)=s^k$.
If one were actually in the region of local duality, and had 
a good approximation to the physical continuum spectral function, then
of course the two methods would be compatible.
Since they are not, 
we conclude that either the OPE is simply not well enough 
converged to provide a reasonable representation of the correlator
away from the timelike real axis, for the $s_0$ values considered
(in which case the whole analysis collapses as a method
for extracting $m_u+m_d$), or the BPR continuum spectral ansatz,
and hence the estimate of $m_u+m_d$ based on it,
is unreliable.  The former seems implausible 
(the contour-improved series, particularly at the somewhat larger
scales shown in the table, appears rather well-behaved) though it cannot
be rigorously ruled out.  

Let us turn now to the SS channel.  Here one replaces
$A^\mu$ above with the $S=-1$ vector current $\bar{s}\gamma^\mu u$,
and obtains sum rules involving the $S=-1$ scalar correlator, $\Pi (s)$.
The analyses of Refs.~\cite{jm,cps} (JM/CPS) and
\cite{cfnp} (CFNP) employ the conventional SVZ
method.  In the former, the Omnes representation, together with experimental
$K_{e3}$ and $K\pi$ phase shift data, is used to fix the
timelike $K\pi$ scalar
form factor at continuum threshold $s=(m_K+m_\pi )^2$.  A sum-of-resonances
ans\"atz for the spectral function, $\rho (s)$,
normalized to this value,
is then employed on the hadronic side.
In contrast, CFNP employ the Omnes representation also above
threshold in order to obtain the $K\pi$ contribution to $\rho (s)$
purely in terms of experimental data.
This improves the low-$s$ behavior of the spectral function (which shows
considerable distortion associated with the attractive $I=1/2$
$s$-wave $K\pi$ interaction).  Unresolved issues for the CFNP analysis
include (1) the 
size of spectral contributions associated with 
neglected higher multiplicity states,
(2) sensitivity to the assumption that the $K\pi$ phase is 
constant at its asymptotic value ($\pi$) beyond the highest
$s$ ($\sim (1.7\ {\rm GeV})^2$), for which it 
is known experimentally, and
(3) the failure to find a stability window
unless the continuum threshold is allowed to lie significantly above
the region of significant $K\pi$ spectral contributions 
(leaving an unphysical region of size $1-2$ GeV$^2$
with essentially no spectral strength in the resulting 
spectral model).
We investigate the CFNP ans\"atz (which retains only the $K\pi$ portion
of $\rho (s)$, obtained as described above)
by studying again
the $w(A,s)$ family of FESR's.  In Table IV are
displayed the OPE and hadronic sides, as a function
of $s_0$ and $A$, for both the JM/CPS and CFNP spectral ansatze.
In each case, the value
of $m_s(1 \ {\rm GeV}^2)$ extracted by the earlier authors
has been used as input on the OPE side.
For the CFNP case, the original authors quote a range of
values; to be specific, we have chosen 
that value from this range, $m_s=155\ {\rm MeV}$,
which produces a match of the OPE and hadronic sides of 
the sum rule for $s_0=4\ {\rm GeV}^2$ and $A=0$.  Comparing the
two sides at other values of $s_0$, $A$ then provides a test of the
quality of the spectral ans\"atz.  The agreement is obviously much better
for the CFNP ans\"atz than 
in the other two cases, though amenable to some further 
improvement.  Note that the most obvious improvement, namely adding
additional spectral strength in the vicinity of the $K_0^*(1950)$
to account for contributions of multiparticle
states (the $K\pi$ branching ratio of the $K_0^*(1950)$ is
$52\pm 8\pm 12\%$\cite{PDG98}),
actually somewhat worsens the agreement between the OPE and
hadronic sides, suggesting that modifications of the spectrum at
lower $s$ may be required.
 
\section{Summary}
We have shown that continuous families
of FESR's can be used
to place constraints on hadronic spectral functions and that,
based both on qualitative physical arguments and a study of the IV channel,
these constraints
can be expected to be more reliable than those based on FESR's
with integer power weights.  The method
is complementary to conventional Borel transformed (SVZ) treatments
in that it involves weights that do not suppress (and
for some $A$ actually enhance) contributions from
the higher $s$ portion of the spectrum.  

Relevant to attempts to extract the light quark masses,
the method has been shown to produce constraints on the continuum
portions of hadronic spectral functions not exposed by previous
sum rule treatments.  For the PI channel, one finds
that either the use of the OPE representation is not 
justified at the scales considered, or existing spectral ansatze
must be modified significantly to produce an acceptable match to the
known OPE representation.  The need for (albeit less significant)
modifications to the CFNP spectral model in the SS channel has also
been illustrated.  We conclude that
the question
of the values of the light quark masses is still open, particularly
for $m_u+m_d$, and that further investigations using
the method explored here may help in  clarifying
the situation.
\acknowledgements
The author acknowledges
the ongoing support of the Natural Sciences and
Engineering Research Council of Canada, and the hospitality of the
Special Research Centre for the Subatomic Structure of Matter at the
University of Adelaide and the T5 and T8 Groups
at Los Alamos National Laboratory, where 
portions of this work were originally performed.
Useful discussions with Tanmoy Bhattacharya and Rajan Gupta,
and with Andreas H\"ocker on the ALEPH spectral
function analysis are also gratefully acknowledged.

\begin{table}
\caption{Integer power weighted OPE and hadronic integrals, $I_k^{ex(OPE)}$
for the IV channel.  All notation as defined
in the text.  Units of $I_k^{ex(OPE)}$ are GeV$^{2(k+1)}$,
and of $s_0$ are GeV$^2$.}\label{Table1}
\vskip .15in\noindent
\begin{tabular}{cccccc}
$s_0$&ex/OPE&$I_0^{ex/OPE}$&\ \ $I_1^{ex/OPE}$&\ \ $I_2^{ex/OPE}$&
\ \ $I_3^{ex/OPE}$\\
\tableline
2.0&ex&.0559$\pm .0006$&.0483$\pm .0007$&.0539$\pm .0010$&.0736$\pm .0017$\\
&OPE&.0612$\pm .0013$&.0584$\pm .0009$&.0746$\pm .0010$&.110$\pm .001$\\
\\
2.2&ex&.0620$\pm .0008$&.0610$\pm .0010$&.0804$\pm .0018$&.129$\pm .004$\\
&OPE&.0667$\pm .0013$&.0704$\pm .0009$&.0998$\pm .0012$&.163$\pm .002$\\
\\
2.4&ex&.0692$\pm .0009$&.0773$\pm .0014$&.117$\pm .003$&.213$\pm .006$\\
&OPE&.0722$\pm .0012$&.0834$\pm .0010$&.130$\pm .001$&.233$\pm .002$\\
\\
2.6&ex&.0765$\pm .0010$&.0953$\pm .0018$&.162$\pm .004$&.324$\pm .009$\\
&OPE&.0777$\pm .0012$&.0976$\pm .0011$&.165$\pm .002$&.322$\pm .003$\\
\\
2.8&ex&.0841$\pm .0013$&.116$\pm .003$&.216$\pm .007$&.470$\pm .018$\\
&OPE&.0832$\pm .0012$&.113$\pm .001$&.207$\pm .002$&.433$\pm .004$\\
\\
3.0&ex&.0924$\pm .0025$&.139$\pm .007$&.285$\pm .019$&.668$\pm .055$\\
&OPE&.0888$\pm .0012$&.129$\pm .001$&.254$\pm .002$&.571$\pm .005$\\
\\
3.1&ex&.0970$\pm .0041$&.154$\pm .012$&.328$\pm .036$&.797$\pm .108$\\
&OPE&.0915$\pm .0012$&.138$\pm .001$&.281$\pm .002$&.651$\pm .005$\\
\end{tabular}
\end{table}

\begin{table}
\caption{Vector isovector channel FESR's for weight functions
with a zero at $s=s_0$.  Columns 2 and 3 contain the hadronic (ex)
and OPE sides, $J_\tau^{ex,OPE}$, of the sum rule 
corresponding to the double zero weight function
occurring in hadronic $\tau$ decay.  Columns 4 and 5, and 6 and 7,
contain the hadronic and OPE sides, $J_{01}^{ex,OPE}$,
$J_{12}^{ex,OPE}$, of the sum rules corresponding
to the weight functions, $w_{01}(s)$, $w_{12}(s)$, 
having a simple zero at $s=s_0$.
All notation is as defined in the text.
Units of $s_0$, $J^{ex(OPE)}_\tau$ and $J_{k-1,k}^{ex(OPE)}$ 
are GeV$^2$.}\label{Table 2}
\vskip .15in\noindent
\begin{tabular}{ccccccc}
$s_0$&$J_{\tau}^{ex}$&$J_{\tau}^{OPE}$&\ \ $J_{01}^{ex}$&$J_{01}^{OPE}$&
\ \ $J_{12}^{ex}$&$J_{12}^{OPE}$\\
\tableline
2.0&.0339$\pm .0005$&.0337$\pm .0010$&.0318$\pm .0005$
&.0323$\pm .0009$&.0107$\pm .0002$&.0103$\pm .0003$\\
2.2&.0364$\pm .0005$&.0363$\pm .0010$&.0343$\pm .0005$
&.0350$\pm .0009$&.0111$\pm .0002$&.0111$\pm .0002$\\
2.4&.0389$\pm .0005$&.0389$\pm .0009$&.0370$\pm .0005$
&.0378$\pm .0008$&.0118$\pm .0002$&.0120$\pm .0002$\\
2.6&.0414$\pm .0005$&.0413$\pm .0009$&.0398$\pm .0005$
&.0406$\pm .0008$&.0127$\pm .0002$&.0128$\pm .0002$\\
2.8&.0440$\pm .0005$&.0440$\pm .0009$&.0428$\pm .0005$
&.0433$\pm .0008$&.0137$\pm .0002$&.0137$\pm .0002$\\
3.0&.0467$\pm .0005$&.0466$\pm .0009$&.0458$\pm .0005$
&.0461$\pm .0008$&.0148$\pm .0002$&.0146$\pm .0002$\\
3.1&.0481$\pm .0005$&.0481$\pm .0009$&.0475$\pm .0005$
&.0475$\pm .0008$&.0154$\pm .0003$&.0150$\pm .0002$\\
\end{tabular}
\end{table}


\begin{table}
\caption{Comparison of the OPE and hadronic (had) sides of the FESR's
corresponding to the weight family $w(A,s)$ defined in
the text, for the pseudoscalar isovector channel.  Units of $s_0$ are
GeV$^2$ and of the $w$-weighted integrals, $10^{-6}\ {\rm GeV}^6$.
The
known pion pole contribution has been subtracted from both sides of the
sum rule.}
\label{Table 3}
\vskip .15in\noindent
\begin{tabular}{cccccc}
$s_0$&$had/OPE$&$A=0$&$A=3$&$A=6$&$A=9$\\
\tableline
2.0&had&2.18&6.76&11.3&15.9\\
&OPE&4.42&14.5&25.6&34.7\\
2.4&had&3.98&11.8&19.5&27.3\\
&OPE&6.94&20.3&33.7&47.0\\
2.8&had&5.68&15.8&25.9&35.9\\
&OPE&9.68&26.6&43.6&60.5\\
3.2&had&7.22&19.0&30.8&42.6\\
&OPE&12.6&33.4&54.3&75.1\\
\end{tabular}
\end{table}

\begin{table}
\caption{Comparison of OPE and hadronic sides of the
$w(A,s)$ family of FESR's  
for the strangeness $S=-1$
scalar channel and for both the
JM/CPS and CFNP ansatze for the hadronic spectral function.
The OPE results are obtained using the 
values of $m_s$ determined in the earlier (SVZ-style) 
analyses.
Units of
$s_0$ are GeV$^2$, and of the hadronic/OPE integrals,
$10^{-3}$ GeV$^6$.}\label{Table 4}
\vskip .1in\noindent
\begin{tabular}{ccccccc}
Case&$s_0$&$had/OPE$&$A=0$&$A=3$&$A=6$&$A=9$\\
\tableline
JM/CPS&3.2&OPE&4.39&9.45&14.5&19.6\\
&&had&5.76&15.5&25.3&35.1\\
&3.4&OPE&4.77&10.3&15.9&21.4\\
&&had&6.37&16.7&27.1&37.5\\
&3.6&OPE&5.15&11.2&17.3&23.3\\
&&had&6.94&17.8&28.7&39.6\\
&3.8&OPE&5.56&12.1&18.7&25.3\\
&&had&7.50&18.9&30.2&41.6\\
&4.0&OPE&5.97&13.1&20.2&27.3\\
&&had&8.05&19.9&31.7&43.5\\
\tableline
CFNP&3.2&OPE&2.51&5.35&8.20&11.0\\
&&had&2.65&6.56&10.5&14.4\\
&3.4&OPE&2.72&5.84&8.96&12.1\\
&&had&2.85&6.90&11.0&15.0\\
&3.6&OPE&2.94&6.34&9.75&13.2\\
&&had&3.04&7.18&11.3&15.5\\
&3.8&OPE&3.16&6.86&10.6&14.3\\
&&had&3.21&7.41&11.6&15.6\\
&4.0&OPE&3.40&7.40&11.4&15.4\\
&&had&3.37&7.40&11.8&16.1\\
\end{tabular}
\end{table}


\begin{references}
\bibitem{svz}M.A. Shifman, A.I. Vainshtein and V.I. Zakharov, Nucl. Phys.
{\bf B147} (1979) 385, 448
\bibitem{rry}L.J. Reinders, H. Rubinstein and S. Yazaki, Phys. Rep. 
{\bf 127} (1985) 1
\bibitem{narisonbook}S. Narison, 
{\it QCD Spectral Sum Rules} (World Scientific,
Singapore, 1989)
\bibitem{leinweber}D. Leinweber, Ann. Phys. (N.Y.) {\bf 254} (1997) 328
\bibitem{shankar}R. Shankar, Phys. Rev. {\bf D15} (1977) 755
\bibitem{mpr77}R.G. Moorhose, M.R. Pennington and G.G. Ross, Nucl.
Phys. {\bf B124} (1977) 285
\bibitem{ck7778}K.G. Chetyrkin and N.V. Krasnikov, Nucl. Phys. {\bf B119}
(1977) 174; K.G. Chetyrkin, N.V. Krasnikov and A.N. Tavkhelidze,
Phys. Lett. {\bf 76B} (1978) 83; N.V. Krasnikov, 
A.A. Pivovarov and N.N. Tavkhelidze,
Z. Phys. {\bf C19} (1983) 301
\bibitem{fndr79}E.G. Floratos, S. Narison and E. de Rafael, Nucl. Phys.
{\bf B155} (1979) 115
\bibitem{bldr85}R.A. Bertlmann, G. Launer and E. de Rafael, Nucl. Phys.
{\bf B250} (1985) 61
\bibitem{bpr}J. Bijnens, J. Prades and E. de Rafael, Phys. Lett.
{\bf B348} (1995) 226
\bibitem{kmtau}K. Maltman, hep-ph/9804298
\bibitem{pp98}A. Pich and J. Prades, hep-ph/9804462
\bibitem{taurefs}Y.S. Tsai, Phys. Rev. {\bf D4} (1971) 2821;
H.B. Thacker and J.J. Sakurai, Phys. Lett. {\bf B36} (1971) 103;
F.J. Gilman and D.H. Miller, Phys. Rev. {\bf D17} (1978) 1846;
F.J. Gilman and S.H. Rhie, Phys. Rev. {\bf D31} (1985) 1066
\bibitem{tauope}E. Braaten, Phys. Rev. Lett. {\bf 60} (1988) 1606;
Phys. Rev. {\bf D39} (1989) 1458;
S. Narison and A. Pich, Phys. Lett. {\bf B211} (1988) 183;
E. Braaten, S. Narison and A. Pich, Nucl. Phys. {\bf B373}
(1992) 581;  F. Le Diberder and A. Pich, Phys. Lett.
{\bf B286} (1992) 147;
K.G. Chetyrkin and A. Kwiatkowski, Z. Phys. {\bf C59}
(1993) 525; S. Narison and A. Pich, Phys. Lett. {\bf B304} (1993) 359;
A. Pich, Nucl. Phys. Proc. Suppl. {\bf 39BC} (1995) 326;
S. Narison, Nucl. Phys. Proc. Suppl. {\bf 40B} (1997) 47
\bibitem{dp92} F. Le Diberder and A. Pich, Phys. Lett. {\bf B289} (1992) 165
\bibitem{pichrev}A. Pich, hep-ph/9704453, to appear in Heavy Flavors II,
eds. A.J. Buras and M. Lindner, World Scientific
\bibitem{ALEPHalphas}A. H\"ocker (for the ALEPH Collaboration), Nucl.
Phys. Proc. Suppl. {\bf 55C} (1997) 379
\bibitem{ALEPH} R. Barate {\it et al.} 
(The ALEPH Collaboration), Z. Phys. {\bf C76} (1997) 15
\bibitem{ALEPH98}R. Barate {\it et al.} (The ALEPH Collaboration),
CERN-EP/98-12, January 1998
\bibitem{pqw}E. Poggio, H. Quinn and S. Weinberg, Phys. Rev. {\bf D13}
(1976) 1958
\bibitem{kniehl96}B.A. Kniehl, Z. Phys. {\bf C72} (1996) 437
\bibitem{pnew}J. Prades, hep-ph/9709405
\bibitem{newqm}F.J. Yndurain, hep-ph/9708300; H.G. Dosch and S. Narison,
Phys. Lett. {\bf B417} (1998) 173; L. Lellouch, E. de Rafael and J. Taron,
Phys. Lett. {\bf B414} (1997) 195
\bibitem{jm}M. Jamin and M. M\"{u}nz, Z. Phys. {\bf C66} (1995) 633
\bibitem{cps}K.G. Chetyrkin, D. Pirjol and K. Schilcher, Phys. Lett.
{\bf B404} (1997) 337
\bibitem{cfnp}P. Colangelo, F. De Fazio, G. Nardulli and N. Paver,
Phys. Lett. {\bf B408} (1997) 340
\bibitem{kmms}K. Maltman, hep-ph/9804299 (in press Phys. Lett. {\bf B})
\bibitem{narison97}S. Narison, hep-ph/9711495
\bibitem{leutwylerqm} H. Leutwyler, Phys. Lett. {\bf B374} (1996) 163;
{\bf B378} (1996) 313; hep-ph/9609467
\bibitem{bgm}T. Bhattacharya, R. Gupta and K. Maltman, Phys. Rev.
{\bf D57} (1998) 5455
\bibitem{gl85}J.~Gasser and H.~Leutwyler, Nucl.\ Phys.\ {\bf B250} (1985) 465
\bibitem{PDG98}Review of Particle Properties, Eur. Phys. J. {\bf C3} (1998) 1
\end{references}
\end{document}